%% file: ms.tex
\documentclass[10pt,a4paper,prx,superscriptaddress,twocolumn,tightenlines]{revtex4-2}

\setcitestyle{numbers,comma,sort&compress,super}
\usepackage{pream} 

\newcites{L}{References}
\newcites{S}{References}
\newcommand\newref[1]{#1}

\newcommand\creflett[2]{\cref{#1}\hyperref[#1]{#2}}
\newcommand\Creflett[2]{\Cref{#1}\hyperref[#1]{#2}}
\newcommand\cw{\text{cw}}
\newcommand\ccw{\text{ccw}}
\newcommand\bs{\text{b}}
\newcommand\pump{\text{p}}
\newcommand\crit{\text{crit}}
\newcommand\trans{\text{trans}}

\begin{document}
{
\linespread{1.0}
\title{Coherent suppression of backscattering in optical microresonators}
\input{authors.tex}

\begin{abstract}
As light propagates along a waveguide, a fraction of the field can be reflected by Rayleigh scatterers. In high-quality-factor whispering-gallery-mode microresonators, this intrinsic backscattering is primarily caused by either surface or bulk material imperfections. For several types of microresonator-based experiments and applications, minimal backscattering in the cavity is of critical importance, and thus, the ability to suppress backscattering is essential. We demonstrate that the introduction of an additional scatterer into the near field of a high-quality-factor microresonator can coherently suppress the amount of backscattering in the microresonator by more than 30~dB. The method relies on controlling the scatterer position such that the intrinsic and scatterer-induced backpropagating fields destructively interfere. This technique is useful in microresonator applications where backscattering is currently limiting the performance of devices, such as ring-laser gyroscopes and dual frequency combs, which both suffer from injection locking. Moreover, these findings are of interest for integrated photonic circuits in which back reflections could negatively impact the stability of laser sources or other components.
\end{abstract}

\maketitle
}

Optical whispering-gallery-mode (WGM) microresonators are widely used in photonics for a range of applications, including sensing and metrology\citeL{Heylman2017, Foreman2015, Righini2011, Kippenberg2011, Matsko2007, DelHaye2007}, optomechanics\citeL{Aspelmeyer2014, Wilson2015, Enzian2019}, quantum optics\citeL{Strekalov2016, Furst2011}, as well as classical and quantum information processing\citeL{MarinPalomo2017, Reimer2016, Pfeifle2014}. Microresonators can adopt a range of different geometries, but for all of them, imperfections in the resonator surface or bulk material can cause scattering of some portion of the light into the counter-propagating whispering-gallery mode\citeL{Mohageg2007, Kippenberg2002, Gorodetsky2000, Weiss1995}.

The backscattered light limits the performance of applications, for example causing unwanted injection locking in laser gyroscopes operating at low rotational speeds\citeL{Chow1985, DellOlio2014, Liang2017}, or in dual frequency combs\citeL{Yang2017, Lucas2018}. Backscattering also reduces the nonlinear enhancement and contributes to back-reflections in devices relying on symmetry breaking of counter-propagating fields\citeL{DelBino2017} for sensing\citeL{Kaplan1981}, optical computing\citeL{Moroney2020, DelBino2020} or isolator\citeL{DelBino2018} applications. Furthermore, control over backscattering permits tuning of the standing wave pattern to maximise coupling by moving an anti-node of the standing wave along the resonator perimeter, which is beneficial for systems relying on evanescent coupling, such as evanescent optomechanics\citeL{Wilson2015}, or biomedical near-field sensors\citeL{Kim2017}. In addition, telecom applications\citeL{Pfeifle2014, Little2000} can benefit from lower backscattering levels.

The imperfections causing backscattering in microresonators are typically distributed around the cavity, but can be approximated as a single scatterer with specific amplitude and phase, as the coherence length of the circulating field is much longer than the cavity round-trip length\citeL{Matres2017}. The elastically reflected field is resonant in the cavity, and therefore builds up in the counter-propagating direction\citeL{MacKintosh1988, Golubentsev1984}. This coupling of the two travelling-wave modes generally results in non-orthogonal chiral eigenmodes composed of unequal superpositions of the two travelling-wave modes\citeL{Mazzei2007}. These superposition modes typically have different frequencies and losses. For high levels of backscattering, the mode splitting may be spectrally resolvable, i.e., detected as two separate resonances\citeL{Kippenberg2002, Weiss1995}. Previous experimental work has focussed on controlling and changing backscattering in such systems with intrinsically high backscattering rates, tuning the mode splitting with a near-field scatterer\citeL{Mazzei2007, Zhu2010} or induced chirality for light flow control\citeL{Peng2016, Wang2020}, without investigating backscattering suppression.
The backscattering problem is now attracting interest in the community, and recently, an optomechanical method to reduce backscattering was demonstrated\citeL{Kim2019}, showing suppression from resolved to unresolved mode splitting.

\begin{figure}[t]
  \centering
  \includegraphics[width=0.5\textwidth]{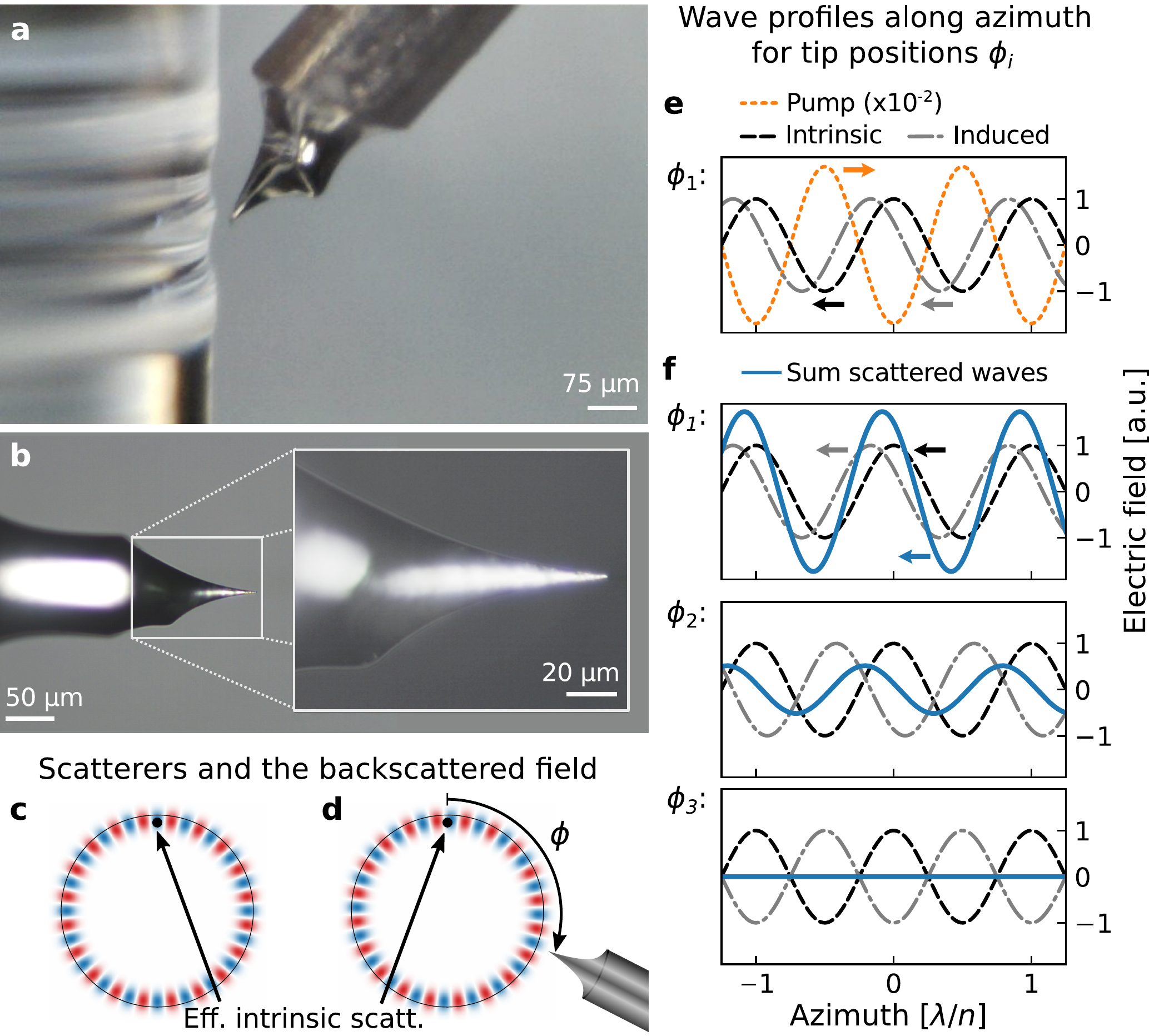}
  \caption{\textbf{Principle of backscattering control.} \textbf{a--b}, Micrographs of the microresonator and tungsten tip. \textbf{c}, Illustration of the real part of the standing wave mode component formed by the intrinsic backscattered field and the pump field (cavity size and wavelength not to scale); the scatterer is here represented by one effective scatterer (black dot). \textbf{d}, Introducing a second scatterer at an azimuthal angular distance $\phi$ from the effective scatterer. \textbf{e}, Illustration of wave profiles along the azimuth showing the counter-clockwise-propagating backscattered waves due to the clockwise-propagating pump field. The backscattered field amplitudes are small compared to the pump as only a small fraction of the light is backscattered. The tip is \textit{critically coupled} (equal amplitudes for intrinsic and induced backscattered waves). \textbf{f}, Backscattered waves and their sum at critical coupling (equal amplitudes) for different azimuthal positions $\phi_i$ of the tip, corresponding to phase offsets between the effective intrinsic scatterer and the induced scatterer $(2m+q)\,\pi$ for integer $m$ and $q = 1/3,\, 5/6,\, 1$, respectively, showing both constructive and destructive interference.}
  \label{fig:principle}
\end{figure}

Here, we show an unprecedented 34 dB suppression of the backscattered light from a WGM resonator, limited by the photodetector noise. This is achieved by manipulating the position of a sub-wavelength-size scatterer within the near field of the optical mode (\creflett{fig:principle}{a}), coherently controlling the effective backscattering. We demonstrate the effect in two silica rod microresonators with intrinsically low backscattering, meaning neither one shows resolved frequency mode splitting (resonator diameters $d = \SI{2.7}{\milli\metre},\,\SI{1.7}{\milli\metre}$ and $Q \simeq \num{2e8},\,\num{1.1e9}$).

In our setup, a sub-wavelength tungsten tip (\creflett{fig:principle}{b-d}) controls the backscattering by coherently scattering additional light from the pumped optical mode into the counter-propagating mode (\creflett{fig:principle}{e}), leading to interference between the intrinsic backscattering and that caused by the metal tip\citeL{Wiersig2011}. With sufficient induced backscattering and an appropriate phase offset between the intrinsic and induced backscatter, the net backscattering can be made to vanish. As the radial position of the tip controls the induced amount of backscattering and the azimuthal position governs the phase offset between the intrinsic and induced backscatter, the tip position can coherently control the net backscattered field (\creflett{fig:principle}{f}). When the tip is inducing backscattering of equal magnitude to the intrinsic backscattering, we call this \textit{critical tip coupling}. The tip also scatters to free-space modes; however, for a small tip diameter, the reduction in the Q-factor is small. This technique enables full control of the amplitude and phase of the backscattering in a microresonator, and in this experiment, we show that it can be reduced by orders of magnitude beyond the unresolved frequency splitting level.

\section{Results}
\subsection{Response of the resonator to the perturbation in the near field.}
\begin{figure}[t]
  \centering
  \includegraphics[width=0.5\textwidth]{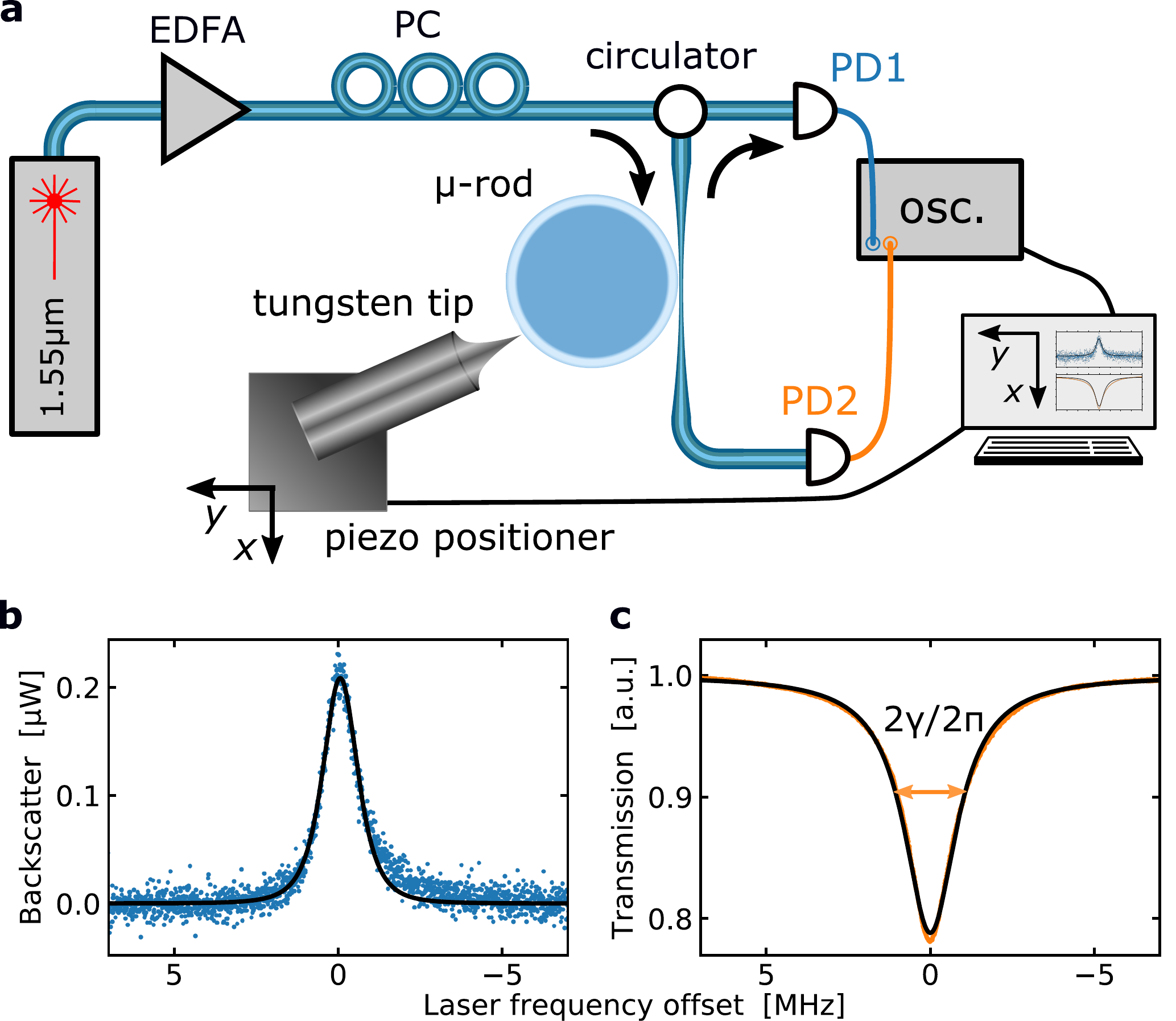}
  \caption{\textbf{Experimental setup and example measurements.} \textbf{a}, Optical circuitry, consisting of a fibre-coupled external cavity diode laser (\SI{1.55}{\micro\meter}) amplified by an erbium-doped fibre amplifier (EDFA), a polarisation controller (PC) to optimise the coupling to the desired resonator mode, a circulator to separate the propagation directions in the fibre, and photodetectors monitor the backscattering (PD1) and transmission of the microresonator (PD2). The tungsten tip near-field probe was fixed to a piezo positioner. Example spectra with fitted Lorentzians (black) of backscattering \textbf{b}, and transmission \textbf{c} for one position of the near-field probe, showing the transmission linewidth $2\gamma$.}
  \label{fig:setup}
\end{figure}

We studied the backscattering amplitude and resonance linewidth as functions of the distance of the tungsten tip from the resonator surface $r$ and its azimuthal position $\phi$, while keeping the tip in the resonator plane ($xy$ plane in \creflett{fig:setup}{a}). The microrod resonator was pumped with a tuneable, \SI{1.55}{\micro\metre} continuous-wave laser using a tapered optical fibre\citeL{Spillane2003} to couple light into the cavity. The $\SI{20}{\milli\watt}$ of input optical power was scanned downwards in frequency in order to obtain spectral data of the resonance. To avoid thermal broadening\citeL{Zhu2019} of the resonance, the laser frequency was scanned at a rate of \SI{450}{\giga\hertz\per\second}. The nonlinear Kerr effect is faster than the scanning rate and thus causes some broadening; however, with low input powers, this broadening is small. A circulator allowed the backscattered light from the cavity to be monitored with a photodetector, along with the cavity transmission (see \creflett{fig:setup}{a}). The field-perturbing probe was fixed to a computer-controlled piezoelectric positioner for three-axis control of the tip position.

\begin{figure*}[t]
  \centering
  \includegraphics[width=0.5\textwidth]{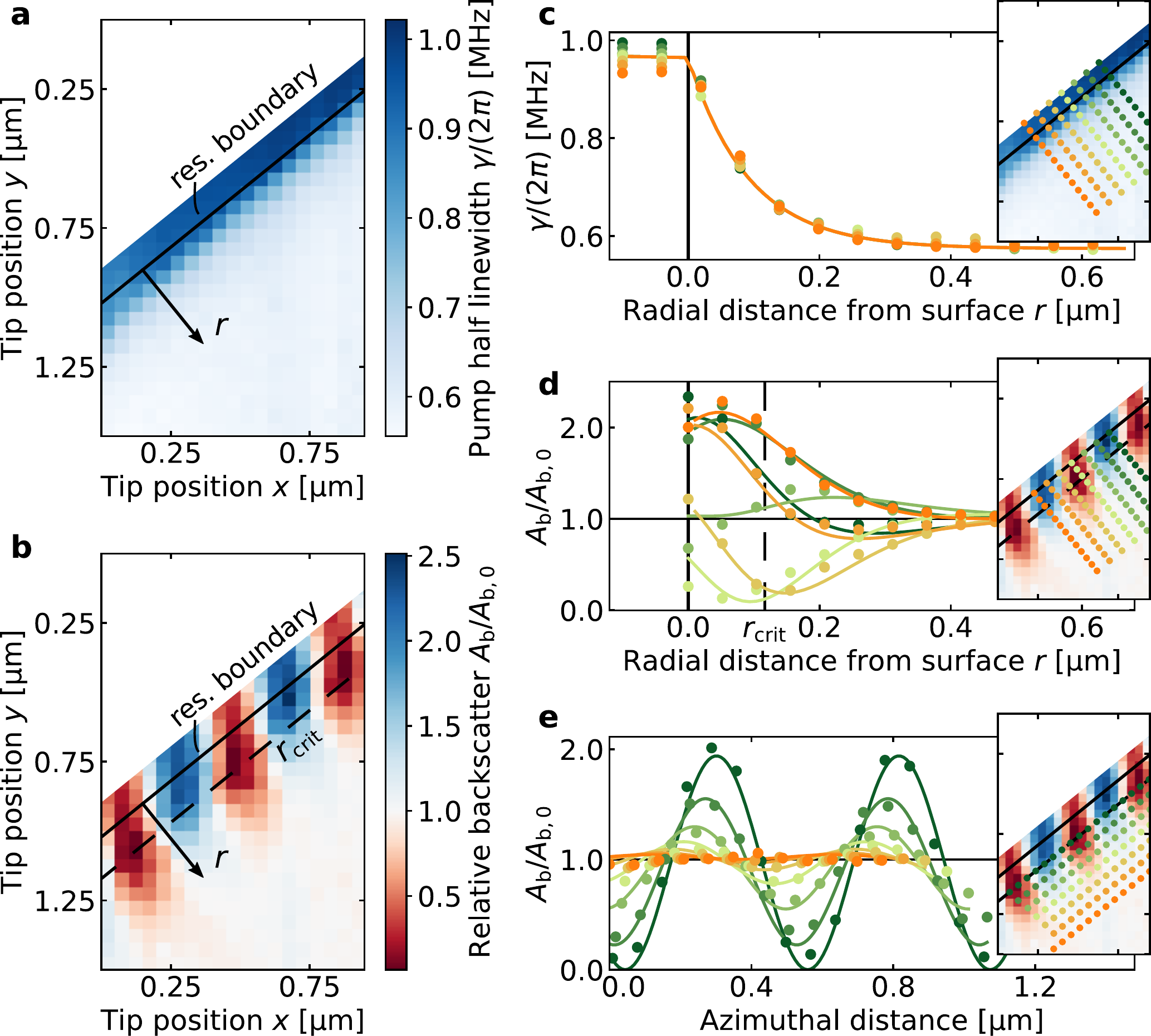}
  \caption{\textbf{Resonator response to the tungsten tip in the near field.} \textbf{a}, Total linewidth (and equivalently $Q$-factor), and \textbf{b}, backscatter amplitude for tip positions in the resonator plane, normalised by the intrinsic backscattering amplitude $A_{\text{b},0}$. The black lines ($r=0$) indicate the fitted resonator surface -- measurements shown as $r<0$ were obtained while the tip was touching the resonator surface, resulting in the tip sliding along the surface. Maximum backscattering suppression is found along the dashed line (\textit{critical tip coupling}); for $r<r_\crit$ the tip is \textit{over-coupled}, reducing the suppression. \textbf{c--e}, Fits (lines) and interpolated data (circles) along cross sections through the experimental data in panels a and b. The insets show where the cross sections are taken. \textbf{c}, Radial tip position dependence of the linewidth and \textbf{d}, the backscattering. \textbf{e},~Azimuthal tip-position dependence of the backscattering.}
  \label{fig:firstmment}
\end{figure*}

The tip position was raster scanned with a step-size of \SI{50}{\nano\metre} over a grid of $(x,y)$ positions in the resonator plane. At each position, transmission and backscattering spectra were simultaneously recorded. The spectra were subsequently fitted with Lorentzian functions (see Materials and methods), as shown for the examples in \creflett{fig:setup}{b} and \hyperref[fig:setup]{c}, to extract the backscatter amplitude $A_\bs$, and the pump-resonance half-linewidth $\gamma$, respectively. The resulting data grids for the resonator with $Q = \num{2e8}$ are shown in \creflett{fig:firstmment}{a} and \hyperref[fig:firstmment]{b}, where each pixel corresponds to a position on the measurement grid. 
For positions corresponding to $r<0$, the tip was touching the resonator and sliding along its surface due to the force applied by the piezo positioner.

We performed numerical fitting of the linewidth and amplitude data vs. the two spatial coordinates to determine the position of the resonator surface and radial distance dependences and estimate the periodicity in the fringe pattern and suppression achieved for the backscattering amplitude.

\subsection{The near-field decay and resonator boundary.}
Assuming a linear coupling between the evanescent near field and the tip, we expect the radial dependence of the linewidth to have the same functional dependence as the energy density in the near field. The evanescent electric field from a waveguide decays exponentially with respect to the perpendicular distance from the surface -- i.e., for a WGM resonator, the radial distance from the surface, $r$. The evanescent field can be expressed $E_\text{ev}(r) = E_\text{surf}\,\exp(-\alpha r)$, where $E_\text{surf}$ is the field strength at the surface, and the decay length is
\begin{equation}
	\label{eq:decay}
	\alpha^{-1} = \frac{\lambda}{2\pi\sqrt{n^2-1}},
\end{equation}
for a field of vacuum wavelength $\lambda$ in a waveguide of refractive index $n$ surrounded by air\citeL{Fornel2001}. With the evanescent field energy density proportional to $|E_\text{ev}(r)|^2 \propto \exp(-2\alpha r)$, we expect the tip-induced loss and backscatter amplitude also to be proportional to this quantity. 

However, the prefactors corresponding to coupling back to the resonator clockwise and counter-clockwise directions, coupling to free-space modes, and absorption by the tip are dependent on the size, geometry, and material of the scatterer, as well as on the polarisation of the mode. As propagating modes of different orders than the mode in question have different resonance frequencies, we do not expect light to couple to other WGM modes. Note that in addition to coupling into free space modes, the tip could also induce losses via coupling into non-guided modes within the fused silica. It has previously been shown that for a silica sub-wavelength tip, it is the tip size relative to the mode volume that determines the amount of induced losses\citeL{Mazzei2007}, and furthermore, the tip can cause mode splitting with no change in the quality factor compared to when the tip is not present\citeL{Gotzinger2002, Wiersig2011}.

The linewidth data were fitted with a piecewise function comprising an exponential decay from the resonator surface, and a linear plateau for the tip positions $r<0$ where it is touching the resonator surface (detailed fitting function given in Materials and methods). The fitted interface between the plateau and the exponential decay determined the resonator surface $r=0$, shown as black lines in \creflett{fig:firstmment}{a} and \hyperref[fig:firstmment]{b}. \Creflett{fig:firstmment}{c} shows cross sections of interpolated values (circles) and the fit (solid) for the linewidth data along the radial direction. The fit gives an exponential decay length for the linewidth of $(2\alpha_\gamma)^{-1} = \SI{92}{\nano\metre}$, compared to the calculated near-field power decay length $(2\alpha)^{-1} = \SI{119}{\nano\metre}$ using \cref{eq:decay} for silica $n=1.44$ at $\lambda=\SI{1553}{\nano\metre}$. The steeper decay coefficient in the experimental data compared to the calculated value can be explained by the tip geometry; as the tip approaches the surface, the effective scattering cross section/polarisability increases, leading to an increasingly larger broadening.

\begin{figure}[t]
	\includegraphics[width=0.42\textwidth]{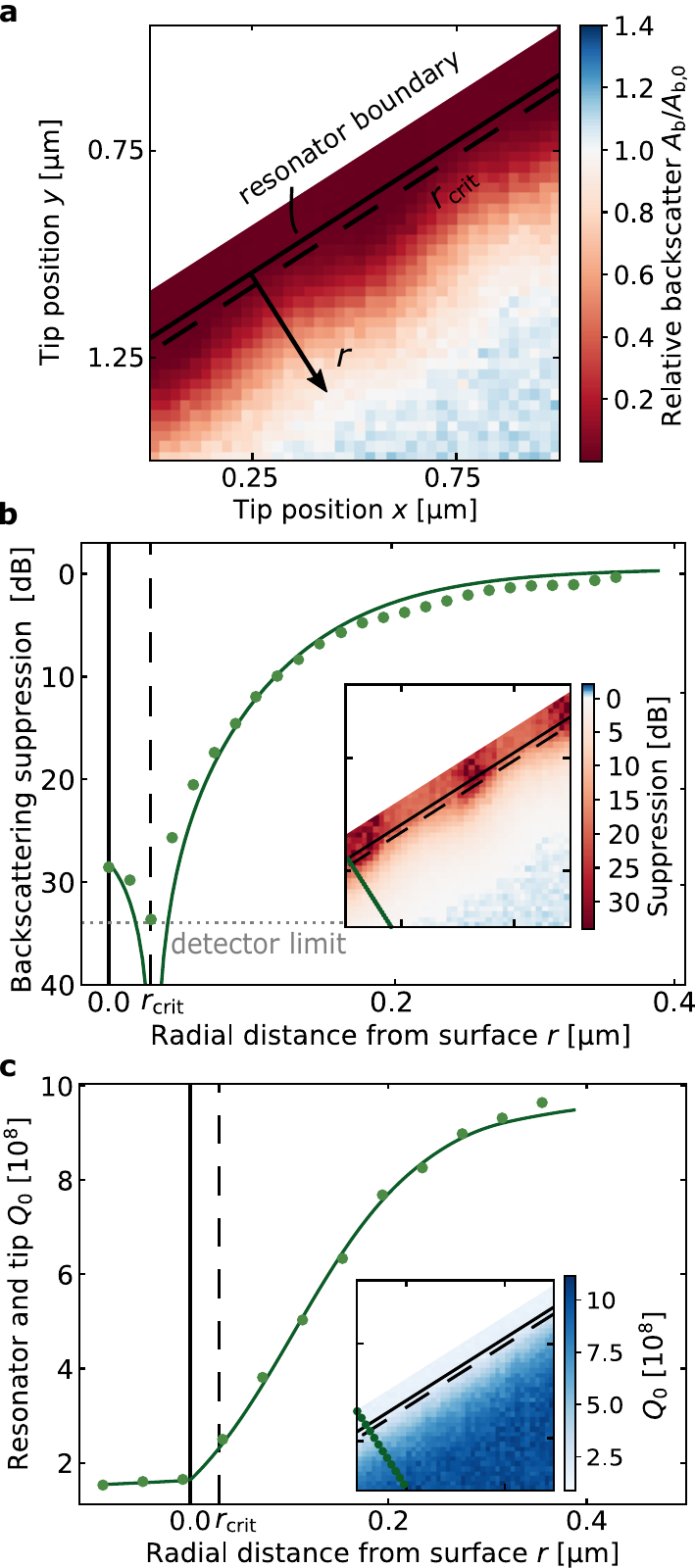}
	\caption{\textbf{Backscattering suppression and $Q_0$ in a high-$Q$ resonator.} \textbf{a}, Backscatter amplitude measurement grid, with resonator boundary and line of `critical tip-coupling' annotated. \textbf{b}, Radial dependence of the backscattering suppression (circles, measurement; line, fit) along a radial line. The noise of the photodetector is indicated by the dotted grey line. Inset: suppression for all positions. The \textit{critical tip-coupling} at which the maximum suppression occurs is annotated. The measured suppression is limited by photodetector noise. \textbf{c}, Radial dependence of resonator and tip quality-factor $Q_0$ along a radial line. Inset: $Q_0$ for all positions.}
	\label{fig:secondmment}
\end{figure}

\subsection{Backscatter suppression analysis.}
The backscattering data in \creflett{fig:firstmment}{b} show maxima and minima when the tip is in the near-field rather than at the surface due to \textit{over-coupling} of the tip -- i.e., the induced backscattering is larger than the intrinsic backscattering. The radial distance at which the tip minimises the backscattering, $r_\crit$, is indicated. The data $r \geq 0$ is fitted with a function which is effectively an exponential decay of coefficient $2\alpha_\bs$, multiplied with a fringe pattern (see Materials and methods). 

The fringe pattern arises due to the relative phase change of the induced backscattering as the tip is translated parallel to the surface (along the azimuthal direction). The expected periodicity of the fringe pattern can be estimated by $\lambda/(2n) = \SI{539}{\nano\metre}$ for silica at our pump wavelength (see Supplementary information). \hyperref[fig:firstmment]{Figures 3d} and \hyperref[fig:firstmment]{e} show cross sections of interpolated values (circles) and fit (solid) along the radial and azimuthal directions, respectively, for the backscattering data. The fit gave a fringe period of $\SI{515}{\nano\metre}$ and a decay coefficient for the backscattering of $(2\alpha_\bs)^{-1} = \SI{99}{\nano\metre}$, similar to the linewidth decay length. The current setup is stable enough for the fringe pattern position to remain in place for measurements of at least 45 minutes.

\subsection{Backscattering suppression in resonators with higher Q-factors.}
\Cref{fig:secondmment} shows data from a similar measurement in a resonator with $Q\simeq \num{1.1e9}$. Due to the extremely high Q-factor, this measurement was performed with a lower scanning speed (\SI{10}{\giga\hertz\per\second}) to avoid ring-down signals, calling for a lower input power ($\sim\SI{40}{\micro\watt}$) to avoid thermal broadening. The backscattering pattern now shows a decrease in the backscattering for all azimuthal positions as the tip approaches the resonator as the backscattering is linewidth dependent (see Methods) and the linewidth now decreases more than for the resonator in \cref{fig:firstmment}. The maximum backscattering suppression obtained, $\SI{34}{\decibel}$, is limited by the photodetector noise.

The total Q-factor of the system was calculated from $Q = \omega_\text{pump}/(2\gamma)$, for pump frequency $\omega_\text{pump} = 2\pi\times\SI{193.1}{\giga\hertz}$. Given the effective taper coupling $\eta$, the quality factor $Q_0$ of the resonator with the tip present can be estimated as $Q_0 = 2Q/(1+\sqrt{1-\eta})$, shown for different radial positions of the tip in \creflett{fig:secondmment}{c}. The maximum suppression occurs when $Q_0 = \num{2.5e8}$.

\section{Discussion}
Optical microresonators provide prospects for miniaturised sensing and communications systems; however, backscattering compromises the performance of some microresonator-based systems. We have demonstrated a method for coherently suppressing the intrinsic backscattering in an optical WGM microresonator, with the suppression exceeding \SI{34}{\decibel} (limited by photodetectors) from a level where frequency splitting is not resolved. Suppression of backscattering opens opportunities for pure travelling-wave resonators, improving the performance in microresonator applications where backscattering is a limiting factor. These applications include symmetry-breaking-based sensing or non-reciprocal systems, optomechanics applications, laser gyroscopes and dual frequency combs; thus, backscattering suppression enables the development of high-accuracy, portable optical spectroscopy systems, gyroscopes and other sensors. The technique is of particular interest for on-chip resonators, where a scatterer can be permanently integrated on chip to coherently suppress back reflections. In addition, one could envision tuneable on-chip backscattering suppression with a MEMS-based device.

{
\small
\input{methods}


\bibliographystyleL{apsrev4-1_asv_titles}
\bibliographyL{refs}


\input{addendum.tex}
}

\onecolumngrid
\clearpage
\newgeometry{left=3cm,right=3cm,top=2cm,bottom=2cm}
\input{supplemental}

\end{document}

%% file: authors.tex
\author{Andreas~\O.~Svela}
\affiliation{National Physical Laboratory, Teddington, TW11 0LW, UK}
\affiliation{Blackett Laboratory, Imperial College London, SW7 2AZ, UK}
\affiliation{Clarendon Laboratory, University of Oxford, OX1 3PU, UK}

\author{Jonathan~M.~Silver}
\affiliation{National Physical Laboratory, Teddington, TW11 0LW, UK}
\affiliation{City, University of London, EC1V 0HB, UK}

\author{Leonardo~Del~Bino}
\affiliation{National Physical Laboratory, Teddington, TW11 0LW, UK}
\affiliation{Heriot-Watt University, Edinburgh, Scotland, EH14 4AS, UK}
\affiliation{Max Planck Institute for the Science of Light, Staudtsta{\ss}e 2, 91058 Erlangen, Germany}

\author{Shuangyou~Zhang}
\affiliation{National Physical Laboratory, Teddington, TW11 0LW, UK}
\affiliation{Max Planck Institute for the Science of Light, Staudtsta{\ss}e 2, 91058 Erlangen, Germany}

\author{Michael~T.~M.~Woodley}
\affiliation{National Physical Laboratory, Teddington, TW11 0LW, UK}
\affiliation{Blackett Laboratory, Imperial College London, SW7 2AZ, UK}
\affiliation{Heriot-Watt University, Edinburgh, Scotland, EH14 4AS, UK}

\author{Michael~R.~Vanner}
\affiliation{Blackett Laboratory, Imperial College London, SW7 2AZ, UK}
\affiliation{Clarendon Laboratory, University of Oxford, OX1 3PU, UK}

\author{Pascal~Del'Haye}
\thanks{pascal.delhaye@mpl.mpg.de}
\affiliation{National Physical Laboratory, Teddington, TW11 0LW, UK}
\affiliation{Max Planck Institute for the Science of Light, Staudtsta{\ss}e 2, 91058 Erlangen, Germany}
\affiliation{Friedrich Alexander University Erlangen-Nuremberg, 91058 Erlangen, Germany}

%% file: methods.tex
\section{Materials and methods}

\subsection{Resonator and tapered fibre fabrication.}
The rod resonators were machined using a \SI{100}{\watt} CO$_2$ laser, milling commercially available 3-\si{\milli\metre}-diameter silica glass rods. The procedure follows that of Del'Haye~\textit{et.~al.}\citeL{DelHaye2013}, but to reach the highest Q-factor of $\num{e9}$, the resonator is fabricated in a nitrogen atmosphere, seeking to avoid the formation of near-IR-absorbing \ce{OH} groups. The rod is fixed to a spindle motor and the glass is evaporated with the focused laser beam. First, the low (high) $Q$ rod was milled down to a 2.7-\si{\milli\metre}-diameter (1.7-\si{\milli\metre}-diameter) cylindrical shape, then the resonator was created by making two ring cuts separated by $\sim\SI{125}{\micro\metre}$ ($\sim\SI{240}{\micro\metre}$).

The tapered fibre was made from a stripped 125-\si{\micro\metre}-diameter standard single-mode silica optical fibre. The fibre was clamped to stepper motor stages, with a hydrogen flame placed under the fibre to heat it while simultaneously pulling symmetrically from both sides.

\subsection{Tungsten tip fabrication.}
The fabrication of the tungsten tip was based on methods used in scanning tunnelling electron microscopy and atomic force microscopy tip fabrication\citeL{Ibe1990,Hagedorn2011}. The process relies surface tension to form a meniscus of solution around a piece of tungsten wire and on the aqueous electrochemical reaction
\begin{equation*}
	\ce{W(solid) + 2 OH^- + 2 H2O -> WO4^2- + 3 H2(gas)}
\end{equation*}
etching the solid tungsten (\ce{W}) anode through oxidation. When the minimum potential difference (\SI{1.43}{\volt}) is overcome, the etching rate varies along the wire due to a hydroxide concentration gradient: the etching rate is slower on the top of the meniscus as the `hydroxide supply' is lower in the meniscus above the horizontal surface. This causes a tip shape to form. Further down, the wire is protected as the tungstate (\ce{WO4^2-}) ions fall along the sides of the wire, forming an increasingly dense laminar layer, protecting the lower end of the tip from being etched. When the diameter at the meniscus is decreased sufficiently, gravity will break off the lower part of the wire.

Temper-annealed, 250-\si{\micro\metre}-diameter, \SI{99.95}{\percent} purity polycrystalline tungsten wire was used in the fabrication. The electrolyte was made by dissolving potassium hydroxide (\ce{KOH}) in deionised water, making a \SI{7.5}{\mol\per\liter} concentration aqueous solution. The second electrode used was tinned copper electrical wire of diameter \SI{0.3}{\milli\metre}. The tungsten wire was pre-etched for five seconds at \SI{4}{\volt} to reduce the surface roughness. After pre-etching, the wire was lifted $\sim\SI{1}{\milli\metre}$ before continuing the etching process at \SI{4}{\volt} until the lower part fell off. The total etching time was approximately two minutes.

\subsection{Experimental setup.}
A fibre-coupled external cavity diode laser at \SI{1.55}{\micro\metre}, connected to an erbium-doped fibre amplifier, was used as the light source in the experiments. To obtain spectral measurements of the microresonator mode, the frequency of the laser source was scanned by current modulation with a triangular wave signal at \SI{1007}{\hertz} (\cref{fig:firstmment}) and \SI{20}{\hertz} (\cref{fig:secondmment}). As the light was subsequently fed into an amplifier operating in saturation, the optical power was kept constant. A polarisation controller was used for optimising coupling to a resonator mode, and a circulator to separate out the backward-propagating light in the tapered fibre for detection. The tapered fibre was mounted on a manual piezo stage to control the coupling to the resonator. Amplified photodetectors and an oscilloscope were used for simultaneously monitoring the transmission and backscattered light.

The tungsten tip was fixed to a polylactic acid (PLA) plastic mount, sitting on a three-axis piezo positioner. The tip position was raster-scanned over a $\SI{1}{\micro\metre}\times\SI{1.5}{\micro\metre}$ area in the resonator plane, and spectra for backscattering and transmission were obtained for each \SI{50}{\nano\metre} (\cref{fig:firstmment}) or \SI{25}{\nano\metre} (\cref{fig:secondmment}) step. The tip-positioner and oscilloscope were simultaneously computer-controlled, allowing a capture time of a few minutes for each of the two measurements reported.

\subsection{Data analysis.}
Least squares fitting procedures were applied to the spectra obtained in the measurement, determining the pump transmission half-linewidth $\gamma$ and amplitude $A_\pump$, and backscattering amplitude $A_\bs$ of the cavity for each position in the measurement. No spectrally resolvable mode splitting was observed.

For the pump resonance, a normal Lorentzian dip from a background $B$ is used, $B-A_\pump/(1+\delta^2/\gamma^2)$, where the detuning with respect to the resonance angular frequency $\omega_0$ is $\delta = \omega - \omega_0$. However, the spectral shape for the backscattering is distorted as it is effectively pumped by a Lorentzian (the pump resonance), resulting in a lineshape $A_\bs/(1+\delta^2/\gamma^2)^2$ in the limit of small backscattering (see Supplementary information for the derivation).

Subsequent to fitting the individual measurements, the grid data of linewidth and backscattering amplitude measurements were fitted. The functions used for the grid data fitting are expressed in a rotated (cartesian) coordinate system $(r,\phi)$ at an angle $\beta$ to the measurement coordinate system $(x,y)$, where the coordinate transformation is given by
\begin{equation*}
	\begin{pmatrix}
		r\\
		\phi
	\end{pmatrix} = 
	\begin{pmatrix}
		\cos\beta & -\sin\beta \\
		\sin\beta &  \cos\beta
	\end{pmatrix}
	\begin{pmatrix}
		x\\
		y
	\end{pmatrix}.
\end{equation*}
In this coordinate system, the $r$ axis is normal to the resonator surface, and $\phi$ can be approximated as the azimuthal position over a short distance compared to the resonator's radius of curvature. The rotation angle $\beta$ was determined as one of the free parameters of the linewidth grid fit, where the linewidth function expressed in the $(r,\phi)$ coordinate system is
\begin{equation*}
	\gamma\,(r,\phi) = \gamma_0 + 
		\begin{cases}
 			a_\pump + s(r-r_0)		& \text{for } r-r_0 < 0 \\
 			a_\pump e^{-2\alpha_\gamma(r-r_0)}	& \text{for } r-r_0 \geq 0
 		\end{cases},
\end{equation*}
with the unperturbed linewidth $\gamma_0$, decay coefficient $\alpha_\gamma$, amplitude of the exponential decay $a_\pump$, linear slope of the plateau at the resonator surface $s$, and the coordinate system offset $r_0$ as free parameters.

The relative backscattering amplitude grid was subsequently fitted with the parameters $\beta$ and $r_0$ fixed to the values obtained from the linewidth fit. Only the portion of data outside the resonator boundary, $r-r_0 = R \geq 0$, was fitted. The function fitted for the backscattering amplitude $A_\bs$ is derived in the Supplementary information, and reads
\begin{equation*}
	A_\bs(R,\phi)
	 = 
		\begin{cases}
			\text{not fitted}		& \text{for } R < 0 \\
 			|g|^2/\gamma^4(r, \phi) 	& \text{for } R \geq 0
 		\end{cases},
\end{equation*}
where $\gamma(r,\phi)$ is the fitted linewidth function, and the coupling from the forward- to the back-propagating mode
\begin{equation*}
	|g|^2 = g_0^2 + 2g_0a_\text{t}\,e^{-2\alpha_\bs R}\cos(\Theta) + a_\text{t}^2\,e^{-4\alpha_\bs R},
\end{equation*}
where $\gamma(r,\phi)$ is the fitted linewidth function, $g_0$ the intrinsic backscattering strength, $\Theta = k_\text{fr}\phi+\theta+\theta_R R$ is a position-dependent phase responsible for the fringe pattern, in which $\theta_R$ is a radially dependent phase accounting for the shape of the tip and/or drift. The period $\Delta$ of the fringe pattern is given by $k_\text{fr}=2\pi/\Delta$.

In order to sample arbitrary lines in the two-dimensional grid of measurement data shown in \creflett{fig:firstmment}{c--e} and \creflett{fig:secondmment}{b--c}, the grid data was interpolated linearly.

%% file: addendum.tex
\subsubsection*{Acknowledgements} This work has been supported by the National Physical Laboratory Strategic Research Programme and the European Research Council (CounterLight, 756966), Marie Sk\l odowska-Curie Actions (MSCA) (CoLiDR, 748519) and UK Research and Innovation (MR/S032924/1). A.\O.S. is supported by an Aker Scholarship and the Engineering and Physical Sciences Research Council (EPSRC) via the Quantum Systems Engineering programme at Imperial College London. J.M.S. acknowledges funding via a Royal Society of Engineering fellowship. L.D.B. and M.T.M.W. are supported by the EPSRC through the Centre for Doctoral Training in Applied Photonics. S.Z. acknowledges funding via an MSCA fellowship (GA-2015-713694).

\newpage
\subsubsection*{Author contributions} A.\O.S. and P.D'H. conceived the experiments. A.\O.S., L.D.B., J.M.S., S.Z. and M.T.M.W. designed and constructed the experiments. 
A.\O.S. performed the experiments and analysed the results with J.M.S. All authors discussed the project, its results, and contributed to the manuscript.

\subsubsection*{Data availability} The data that supports this Article is available from the corresponding author upon reasonable request.

\subsubsection*{Competing interests} The authors declare that they have no competing financial interests.

\subsubsection*{Additional information} Correspondence and requests for materials should be addressed to P.D'H.~(email: pascal.delhaye@mpl.mpg.de).

%% file: supplemental.tex
\setcounter{equation}{0}
\def\theequation{S\arabic{equation}}

\section{Supplementary information}

\subsection{The pump and backscatter lineshapes.}
Following a time-dependent approach\citeS{Rabus2007, Wiersig2011, Woodley2018}, the steady-state equations of motion for the two circulating, counter-propagating fields $e_{\cw,\ccw}$, both detuned by $\delta$ from the pump and perturbed by complex scattering coefficients $g_{jk}$, can be expressed as
\begin{equation*}
	\begin{pmatrix}
		\dot{e}_\cw\\
		\dot{e}_\ccw
	\end{pmatrix} =
	\begin{pmatrix}
		-\gamma - i\delta + ig_{11} & ig_{12}\\
		ig_{21} & -\gamma - i\delta + ig_{22}
	\end{pmatrix}
	\begin{pmatrix}
		e_\cw\\
		e_\ccw
	\end{pmatrix} +
	\begin{pmatrix}
		E_\cw\\
		0
	\end{pmatrix} = 0
\end{equation*}
when a field $E_\cw$ is pumping the clockwise propagating mode. Inverting the matrix, we obtain
\begin{equation*}
	\begin{pmatrix}
		e_\cw\\
		e_\ccw
	\end{pmatrix} =
	\frac{E_\cw}{(\gamma+i\delta - ig_{11})(\gamma+i\delta - ig_{22}) + g_{12}g_{21}}\begin{pmatrix}
		\gamma + i\delta - ig_{22}\\
		ig_{21}
	\end{pmatrix}.
\end{equation*}
In the small-backscattering regime $|g_{jk}| \ll \gamma$, this gives
\begin{align}
	\label{eq:fields}
	e_\cw = \frac{E_\cw}{\gamma +i\delta}; &&	e_\ccw = \frac{ig_{21}E_\cw}{(\gamma +i\delta)^2},
\end{align}
where $|e_\cw|^2, |e_\ccw|^2$ are proportional to the powers circulating in the respective directions.

The output fields in the taper from the cw and ccw directions can be expressed\citeS{Rabus2007} using the taper coupling linewidth $\kappa$,
 \begin{align*}
	E_{\cw,\,\trans} = E_\cw - 2\kappa e_\cw; &&
	E_{\ccw,\,\trans} = E_\ccw - 2\kappa e_\ccw = - 2\kappa e_\ccw.
\end{align*}
To find the lineshapes of the output fields, insert \cref{eq:fields} and take the modulus squared to obtain
 \begin{align*}
	|E_{\cw,\,\trans}|^2 = |E_\cw|^2 \left(1 - \frac{4\kappa(\gamma - \kappa)}{\gamma^2 + \delta^2}\right); &&
	|E_{\ccw,\,\trans}|^2 = |g_{21}|^2|E_\cw|^2 \frac{4\kappa^2}{(\gamma^2 + \delta^2)^2}.
\end{align*}
The total linewidth has two components, the intrinsic losses $\gamma_0$ and the taper coupling $\kappa$, such that $\gamma = \gamma_0 + \kappa$. Furthermore, the coupling efficiency $\eta = 4\kappa\gamma_0/\gamma^2$, giving
 \begin{align}
 	\label{eq:lineshapes}
	|E_{\cw,\,\trans}|^2 = |E_\cw|^2 \left(1 - \frac{\eta}{1 + \delta^2/\gamma^2}\right); &&
	|E_{\ccw,\,\trans}|^2 = |g_{21}|^2|E_\cw|^2 \frac{4\kappa^2/\gamma^4}{(1 + \delta^2/\gamma^2)^2}.
\end{align}
This shows that the transmitted clockwise pump has a dip with a normal Lorentzian lineshape, whereas the backscattered power will exhibit a peak with a squared Lorentzian lineshape of amplitude $A_\bs \propto  |g_{21}|^2/\gamma^4$.

\subsection{Derivation of the backscattering fitting function.}

From \cref{eq:lineshapes}, we are expecting the backscattering power at resonance to be proportional to $|g_{21}|^2/\gamma^4$. The coupling coefficient $g_{21}$ has two contributions, an intrinsic $g_0$, which by a suitable choice of the relative phase between the cw and ccw basis states can be made to be real, and a tip-induced $g_\text{tip}$. The tip-induced coupling is expected to follow $g_\text{tip} = a_\text{t}\,e^{-2\alpha_\bs R}e^{i\Theta}$, where $\Theta = k_\text{fr}\phi+\theta+\theta_R R$ is a position-dependent phase responsible for the fringe pattern, in which $\theta_R$ is a radially dependent phase accounting for the shape of the tip and/or drift. Coherently adding the two contributions, $|g_0 + g_\text{tip}|^2$, we obtain
\begin{equation*}
	|g_{21}|^2 = g_0^2 + 2g_0a_\text{t}\,e^{-2\alpha_\bs R}\cos(\Theta) + a_\text{t}^2\,e^{-4\alpha_\bs R}.
\end{equation*}
As only the portion of data outside the resonator boundary, $r-r_0 = R \geq 0$, was fitted, the fitting function for the backscattering amplitude can be expressed as
\begin{equation}
	\label{eq:bs}
	A_\bs(R,\phi)
	 =
		\begin{cases}
			\text{not fitted}		& \text{for } R < 0 \\
 			\frac{\displaystyle g_0^2 + 2g_0a_\text{t}\,e^{-2\alpha_\bs R}\cos\left[\Theta(R,\phi)\right] + a_\text{t}^2\,e^{-4\alpha_\bs R}}{\displaystyle \gamma^4(r, \phi)} 	& \text{for } R \geq 0
 		\end{cases}\,,
\end{equation}
in which $\gamma(r,\phi)$ is the fitted linewidth function.

\subsection*{Expected period of the fringe pattern.}

The fringe pattern in the backscattering arises from the $\cos(\Theta)$ term in \cref{eq:bs}, where $\Theta(R,\phi)$ is the phase of the backscattered light from the tip. We find the periodicity $\Delta$ of the fringe pattern with respect to the azimuthal position $\phi$ by considering the distance between subsequent maxima. As the phase of the backscattered light depends on the phase of the clockwise mode relative to the ccw mode at the position of each part of the tip involved in backscattering, this relative phase difference varies with $\phi$ by $2k_\text{opt}\phi$, where $k_\text{opt}$ is the optical wavenumber. Hence, the distance $\Delta$ between maxima is given by $\Theta_{m+1}-\Theta_m = 2\pi = 2k_\text{opt}(\phi_{m+1} - \phi_m) =2k_\text{opt}\Delta$.

For a resonator of material with refractive index $n$, the optical wavenumber $k_\text{opt} \simeq 2\pi n/\lambda$ for the vacuum wavelength $\lambda$ (the approximation is due to the transverse confinement of the mode), giving a period $\Delta \simeq \lambda/(2n)$.

{
\small\linespread{1.0}
\bibliographystyleS{apsrev4-1_asv_titles}
\bibliographyS{refs}
}